\documentstyle[epsfig]{jaa}

\begin{document}

\title[TreePM Method for Two-Dimensional Cosmological Simulations]
{TreePM Code for Two-Dimensional Cosmological Simulations}
\author[Suryadeep Ray]{Suryadeep Ray \\
Harish-Chandra Research Institute \\
Chhatnag Road, Jhunsi \\
Allahabad-211019, India. \\
e-mail : surya@mri.ernet.in}

\maketitle
\label{firstpage}

\begin{abstract}
We describe the two-dimensional TreePM method in this paper. 
The 2d TreePM code is an accurate and efficient technique to 
carry out large two-dimensional N-body simulations in cosmology. 
This hybrid code combines the 2d Barnes and Hut Tree method 
and the 2d Particle-Mesh method. We describe the splitting 
of force between the PM and the Tree parts. We also estimate 
error in force for a realistic configuration. Finally, we discuss
some tests of the code.
\end{abstract}


\section{Introduction}

It is believed that large-scale structures in the Universe 
have formed from the gravitational amplification of initial seed 
density perturbations. Evolution of density perturbations at 
scales smaller than the Hubble radius in an expanding background 
can be studied in the Newtonian limit in the matter-dominated 
regime. Linear theory can be used to study the growth of small 
perturbations in density. But in the absence of analytical methods, 
numerical simulations are the only tool available for studying 
clustering in the non-linear regime. The last two decades have 
seen a rapid development of techniques of cosmological simulations 
as well as computing power and the results of these simulations have 
provided valuable insight into the study of structure formation.  

A number of attempts has been made over the past decade 
to model the non-linear evolution of constructs like the two-point
correlation function using certain non-linear scaling relations
\cite{Hamilton 1991}, \cite{Nityananda 1994}. In these relations, the
evolution of the correlation function can be divided into three
distinct regimes \cite{Padmanabhan 1996} - the linear regime, 
the intermediate regime and the non-linear regime. Clearly, a 
large dynamic range is required in any N-body simulation in order 
to address the issue in all the three regimes under consideration.
It has been pointed out \cite{Bagla 1998}, \cite{Munshi 1998} 
that by simulating a two-dimensional system a much higher dynamic 
range can be achieved as compared to a complete three-dimensional 
simulation with similar computational resources. 

The simplest N-Body method that has been used for studying clustering
of large scale structure is the Particle Mesh method. It has two elegant
features in that it provides periodic boundary conditions by default,
and the force is softened naturally so as to ensure collisionless
evolution of the particle distribution.  However, softening of force
done at grid scale implies that the force resolution is very poor.
In particular, this limits the dynamic range over which we can trust the
results of the code \cite{Bouchet 1985}, \cite{Bagla 1997}. 

A completely different approach to the problem of computing force is
used in the Tree method. In this approach we consider groups of
particles at a large distance to be a single entity and compute the
force due to the group rather than sum over individual particles. There
are different ways of defining a group, but by far the most popular 
method is that due to Barnes and Hut \cite{Barnes 1986}. 

Several attempts have been made to combine the high resolution of a 
Tree code with the natural inclusion of periodic boundary conditions 
in a PM code \cite{Xu 1995, Bode 2003, Dubinsky 2004}. The TreePM 
method in three dimensions is a hybrid N-body method which attempts 
to combine the same features \cite{Bagla 2002}. The basic motivation 
for these codes is to improve the acceptable dynamic range of 
simulations without a proportionate increase in computational 
requirements. In this paper, we describe the TreePM method for a 
two-dimensional system.

The plan of the paper is as follows: \S{2} introduces the basic
formalism of both Tree and Particle-Mesh codes. \S{3} describes the 
modelling of the force in two dimensions. \S{4} gives the mathematical 
model for splitting the force in two dimensions between the Tree force
and the Particle-Mesh force components. \S{5} describes the
softening scheme used for the 2d force and we analyse errors 
in force for the 2d TreePM code in \S{6}. We discuss
the integration of the equations of motion that we use in the 
2d TreePM code in \S{7}. We also describe a test for self-similar 
evolution of power law spectra in the same section. We present some  
results of a 2d TreePM simulation run in \S{8}. A discussion of the 
relative merits of the TreePM code and a PM code is also given. 
Computational requirements of our implementation of the 2d TreePM 
code are discussed in \S{9}.   

\section{A Review of the Tree and the Particle-Mesh Methods}

\subsection{The Tree Method}

We use the same method as the Barnes and Hut (1986) Tree code in our 
implementation of the 2d TreePM code. In a 2d Tree code the simulation 
area is taken to be a square. If this were to represent the stem of a 
tree, then it will be subdivided at each stage into smaller squares 
(branches) till we reach the particles (leaves). To construct the 
tree we add particles to the simulation area and subdivide any 
cell that ends up with two particles \cite{Barnes 1986}.

The force on a particle is computed by adding contribution of other
particles or of cells.  If a cell is too close to the particle, or if it
is too big, we consider the sub cells of the cell in question instead.
The decision is made by computing the quantity $\theta$ and comparing it
with a threshold $\theta_c$:
\begin{equation}
\theta = \frac{d}{r} \leq \theta_c  \label{trwalk}
\end{equation}
where $d$ is the size of the cell and $r$ is the distance from the
particle to the centre of mass of the cell.  The error in
force increases with $\theta_c$.

The number of terms that contribute to the force of a particle is 
much smaller than the total number of particles for most choices 
of $\theta_c$ and this is where a tree code gains on a direct force 
summation method. 
 
We will use the Barnes and Hut (1986) Tree code, as already
mentioned. A crucial change to the standard tree walk is that we 
do not follow nodes representing cells that do not have any spatial 
overlap with the region within the threshold radius ($r_{cut}$, 
defined later) for computing the short range force. 

\subsection{The Particle-Mesh Method}

A Particle-Mesh (PM) code is the obvious choice for computing
long range interactions. A PM code adds the construct of a 
regular grid to the distribution of particles. The density field 
represented by particles is interpolated onto grid points 
and the Poisson equation is solved in Fourier space. The force is 
then interpolated back to the positions of particles. Use of a grid 
makes forces inaccurate at the grid scale and smaller scales. In 
this scheme, the mesh and the weight function 
(Cloud-in-Cell (CIC) in our case) used for interpolation 
between the grid and particle positions are the main sources of
anisotropy. However, we use the Particle-Mesh method only for 
computing the long range force and errors at small scales do not 
contribute significantly. Also, by {\it{de-convolving}} the 
interpolating function \cite{Bagla 2002a}, we reduce errors due to 
anisotropy effects substantially. 

\section{The Gravitational Force in Two Dimensions}

When we go from three to two dimensions, we have, in principle, two
different ways of modelling the system \cite{Bagla 1998} : (1) 
We can consider two-dimensional perturbations in a three-dimensional 
expanding Universe. Here we take the force between particles to 
be $1 \over r^2$ and assume that all particles (representing 
perturbations) and their velocities are confined to a single plane at the 
initial instant. (2) We can study perturbations that do not depend on one 
of the three coordinates, i.e., we start with a set of infinitely long 
straight ``needles'' all pointing along one axis. The force of 
interaction then falls as $1 \over r$. The evolution keeps the 
``needles'' pointed in the same direction, and we study the clustering in 
an orthogonal plane. Particles in the N-body simulation represent the 
intersection of these ``needles''with this plane. In both of these 
approaches, the Universe is three-dimensional and the background is 
expanding isotropically. Following earlier studies \cite{Bagla 
1998},\cite{Filmore 1984}, \cite{Munshi 1998}, we choose the second 
of the two options.

More specifically, in order to obtain the force due to perturbations
in a plane, we solve the Poisson equation for the perturbed part of the 
gravitational potential in two dimensions, whereas the unperturbed 
background is still the three dimensional {\it{spherically symmetric}}
Friedman Universe. Thus the perturbations are described by the {\it{mass 
per unit length}}, where this length is in the direction orthogonal to 
the two dimensions considered here.

The gravitational force of a particle situated at the origin in two
dimensions then has the form : 
\begin{equation}
{\bf{f}}({\bf{r}}) = - \left[{G m \over r^2}\right] {\bf{r}} .
\label{2dforce}
\end{equation}
Here $G$ is the gravitational coupling constant and $m$ is the mass
per unit length of the ``needle'' represented by the particle.

\section{The Mathematical Model for the 2d TreePM Code}

We split the $1 \over r$ force into a long range force and a short 
range force in a manner identical to that for the three dimensional TreePM 
force \cite{Bagla 2002}. We compute the long range force in Fourier 
space and the short range force in real space. Following Ewald's method 
\cite{Ewald 1921}, the gravitational potential can be split into two parts 
in Fourier space :
\begin{eqnarray}
\phi_k &=& - \frac{2 \pi G \rho_k}{k^2} 
\label{pm_std} \\
&=& - \frac{2 \pi G \rho_k}{k^2} \exp\left(-k^2 r_s^2\right)  -
 \frac{2 \pi G \rho_k}{k^2} \left[1 - \exp\left(-k^2
 r_s^2\right)\right]\nonumber\\ 
 &=& \phi_k^l + \phi_k^s \nonumber \\ 
\phi_k^l &=& - \frac{2 \pi G \rho_k}{k^2} \exp\left(-k^2
 r_s^2\right) 
\label{longr} \\
\phi_k^s &=& - \frac{2 \pi G \rho_k}{k^2} \left[1 - \exp\left(-k^2
 r_s^2\right)\right] 
\label{shortr}
\end{eqnarray}
where $\phi^l$ and $\phi^s$ are the long range and the short range
potentials, respectively. The splitting is done at the scale $r_s$. 

The expression for the 2d short range force in real space is 
\begin{equation}
{\bf f}^s({\bf r}) = - \exp\left[-\frac{r^2}
{4 {r_s}^2}\right]\frac{G m}{r^2}{\bf r} 
\label{fshort2d}
\end{equation}
The above equations describe the mathematical model for force in the
2d TreePM code. The long range potential is computed in Fourier
space, just as in a PM code, but using eqn.\ref{longr} instead of
eqn.\ref{pm_std}. This potential is then used to compute the long
range force. The short range force is computed directly in real space
using eqn.\ref{fshort2d}. This is computed using the Tree approximation. 
The short range force falls rapidly at scales $r \gg r_s$ and 
hence we need to take this into account only in a small region 
around each particle. We call the scale upto which we add the small 
scale force as $r_{cut}$.  

Evaluation of the short range force can be time consuming. To save 
time, we compute an array containing the magnitude of the short range 
force at the outset. This procedure is identical to that followed 
in the 3d TreePM code \cite{Bagla 2002}. The force between any two 
objects, particle-cell or particle-particle, is then computed by 
linearly interpolating between the nearby array elements multiplied 
by the unit vector ${\bf r}$. It is necessary for the array to 
sample the force at sufficiently closely spaced values of $r$ in 
order to keep interpolation errors in control.

\section{Softening of the Force}

We need to soften the 2d gravitational force at small scales in order to 
ensure {\it{collisionless evolution}} of the particle distribution in a 
cosmological simulation. We have considered two schemes for softening 
of the force at small scales :

1) {\bf{Plummer Softening}}. 
The force, in this case, will be given by
\begin{equation}
{\bf{f}}({\bf{r}}) = - {1 \over (r^2 + \epsilon^2)} {\bf{r}}
\label{Plummerforce}
\end{equation}
where $\epsilon$ is the softening length.

2) {\bf{Cubic Spline Softening}}. 
In this case, we solve the Poisson equation in two dimensions 
with the force due to a point mass replaced by that exerted by 
an extended mass distribution represented by the following :
\begin{equation}
\rho(r) = m W(r,\epsilon)
\label{cubicspline_source}
\end{equation} 
Here $W(r,\epsilon)$ is the normalised spline kernel used in the 
SPH formalism \cite{Monaghan 1992} with $\epsilon$ the smoothing 
length. 

$W(r,\epsilon)$ has the following form in two dimensions.
\begin{equation}
W(r,\epsilon) = \left({40 \over 7 \pi \epsilon^2}\right) 
\left\{ \begin{array}{ll}
1 - 6 \left({r \over \epsilon}\right)^2 + 6 \left({r \over \epsilon}\right)^3,
& 0 \geq {r \over \epsilon} \leq 0.5 \\
2 \left(1 - {r \over \epsilon}\right)^3, & 0.5 < {r \over \epsilon} 
\leq 1.0 \\
0 & {r \over \epsilon} > 1.0 \end{array} \right. 
\label{splinekernel}
\end{equation}

Solving the Poisson equation and using relevant boundary conditions for 
the potential and its first derivative (i.e. the force) to obtain the 
constants of integration, we get the cubic spline softened potential and 
then obtain the force $\bf{f}$ as a gradient of the potential.
\begin{equation}
{\bf{f}}({\bf{r}}) = \left\{ \begin{array}{ll}
- \left\{{10} - {30 r^2 \over \epsilon^2}
+ {96 r^3 \over 5 \epsilon^3} \right\} 
{4 G m \over 7 \epsilon^2} {\bf{r}}   
& 0 \geq {r \over \epsilon} \leq 0.5 ; \\ 
- \left\{{80 \over \epsilon^2} - 
{160 r \over \epsilon^3}
+ {120 r^2 \over \epsilon^4}
- {32 r^3 \over \epsilon^5}
- {1 \over r^2}\right\} {G m \over 7} {\bf{r}} 
& 0.5 < {r \over \epsilon} \leq 1.0 ; \\ 
- {G m \over r^2} {\bf{r}} & {r \over \epsilon} > 1.0 \end{array} \right.
\label{2dforcesoft}
\end{equation}

One can easily see that the spline softening has the advantage that the 
force becomes exactly Newtonian for $r > \epsilon$, while the Plummer 
force converges relatively slowly to the Newtonian form. Dehnen 
\cite{Dehnen 2001} has argued that compact softening kernels are
superior and we use the spline softened kernel in our implemention 
of the 2d TreePM code.

\section{Error Estimation}

It is important to estimate the errors in numerical evaluation of force 
in a realistic situation, even though we do not expect errors to add up 
coherently. For a comprehensive study in errors in force 
introduced by various components of the TreePM code we would like to refer to 
\cite{Bagla 2002a}. Though the above-mentioned work is in the context of
the 3d TreePM code, the key features of the analysis actually carry 
over to the 2d TreePM code as well.

We test errors for two distributions of particles : a
homogeneous distribution and a clumpy distribution. For the
homogeneous distribution, we use randomly distributed particles 
in a box. We use $1024^2$ particles distributed on a $1024^2$ 
grid. We compute the force using a reference setup 
($r_s=4$, $\theta_c=0.01$, $r_{cut}=6 r_s$) and the setup we wish to 
test ($r_s=1$, $\theta_c=0.5$, $r_{cut}=5 r_s$). We compute the 
fractional error in force acting on each particle. This is defined as 
\begin{equation}
\epsilon = \frac{\left\vert {\bf f} - {\bf f}_{ref}
\right\vert}{\left\vert {\bf f}_{ref} \right\vert}  .
\end{equation}

\begin{figure}
\epsfxsize=4truein\epsfbox[38 27 513 506]{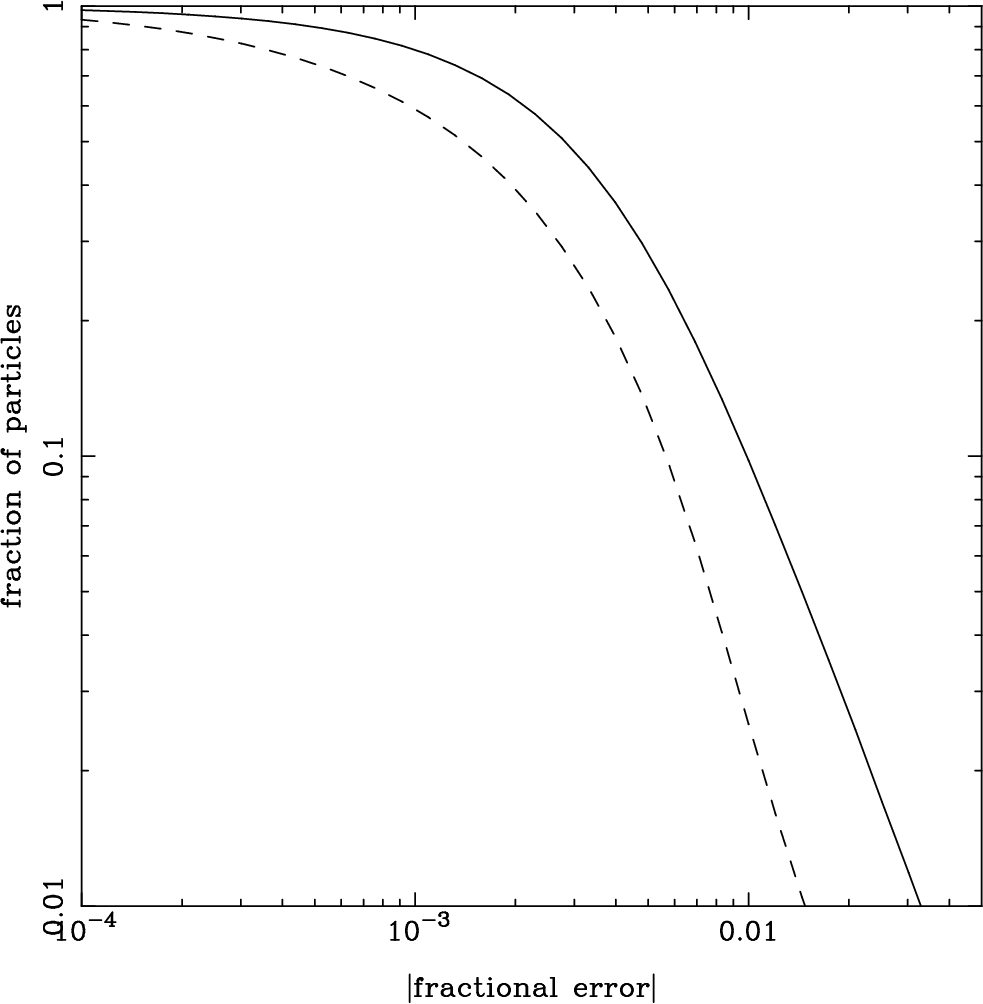}
\caption{This figure shows the distribution of errors. The variation
of the fraction of particles with error greater than a threshold, as a
function of the threshold error, is plotted. Thick line marks
the error for a homogeneous distribution of particles and the dashed
line shows the same for a clumpy distribution. These errors were
measured with respect to a reference force, determined with a very
conservative value of $r_s$ and $\theta_c$. This panel shows that 
$99\%$ of the particles have fractional error in force that
is less than $4\%$ for the homogeneous distribution and less than
$2\%$ for the clumpy distribution.}
\label{error_plot}
\end{figure}

Fig.\ref{error_plot} shows the cumulative distribution of fractional 
errors. The curves show the fraction of particles with error greater 
than $\epsilon$. The thick line shows this for the homogeneous
distribution. Error $\epsilon$ for $99\%$ of particles is less than
$4\%$. Results for the clumpy distribution of particles are shown
by the dashed line. For this case, we used the output of a 2d power 
law ($n=1$) simulation run in an Einstein deSitter background Universe 
with the TreePM code. Errors in this case are somewhat smaller as 
compared to the homogeneous distribution for much the same reason as 
that for a 3d Tree code \cite{Hernquist 1991} or a 3d TreePM code 
\cite{Bagla 2002}. Error $\epsilon$ for $99\%$ of particles is less 
than $2\%$ for the clumpy distribution.

\section{Integrating the Equation of Motion}

Our discussion so far has dealt only with the evaluation of force.
This is the main focus of this paper as the key difference between the
TreePM and other methods is in the scheme used for evaluation of
force.  However for the sake of completeness, we give here details of
integration of the equations of motion used in the code. We use 
an Einstein deSitter background cosmology for all our 2d simulations. 
The equations of motion are then given by the simple form
\begin{eqnarray}
\ddot{\mathbf x} + 2 \frac{\dot{a}}{a} \dot{\mathbf x} &=& -
\frac{1}{a^2} \vec\nabla\phi
\nonumber\\
\nabla^2\phi &=& 4 \pi G a^2 \left(\rho - {\bar\rho} \right)
\label{eqnmot}
\end{eqnarray}
Here ${\mathbf x}$ is the comoving coordinate, $a$ is the scale
factor, $\phi$ is the gravitational potential of perturbations, 
$\rho$ is the total density and ${\bar\rho}$ is the average density 
of the Universe. Dot represents differentiation with respect to time.
We can recast these equations in the following form.
\begin{eqnarray}
{\mathbf x}'' +  \frac{3}{2 a} {\mathbf x}' 
&=& - \frac{3}{2 a} \vec\nabla\psi
\nonumber \\
\nabla^2\psi &=& \delta = \frac{\rho}{\bar\rho} - 1 
\label{eqnmot_treepm}
\end{eqnarray}
Here prime denotes differentiation with respect to the scale factor,
$\delta$ is the density contrast and $\psi$ is the appropriately
scaled gravitational potential of perturbations.

The equations of motion are identical to that in three dimensions 
apart from the fact that all the vectors under consideration are 
two-dimensional vectors. The functional form of the gravitational 
force given by ${\vec{\nabla}} \phi$ is, of course, different. 

One can see that the equations of motion contain a velocity dependent 
term and hence we cannot use the usual leap-frog method. We recast the 
leap-frog method so that velocities and positions are defined at the 
same instant \cite{Bagla 2002a}. We solve the equation for velocity 
iteratively. Time step is chosen to be a small fraction of the smallest 
dynamical time in the system at any given stage. The fraction to be chosen 
is fixed by checking for {\it{scale invariance}} in evolution of 
power law spectra : a simulation is repeated with different choices of 
timestep until we find the largest timestep for which we can reach the 
highly non-linear regime and retain scale invariance as well. We then 
use a timestep that is half of this largest time step. 

\begin{figure}
\epsfysize=4truein\epsfbox[10 28 524 506]{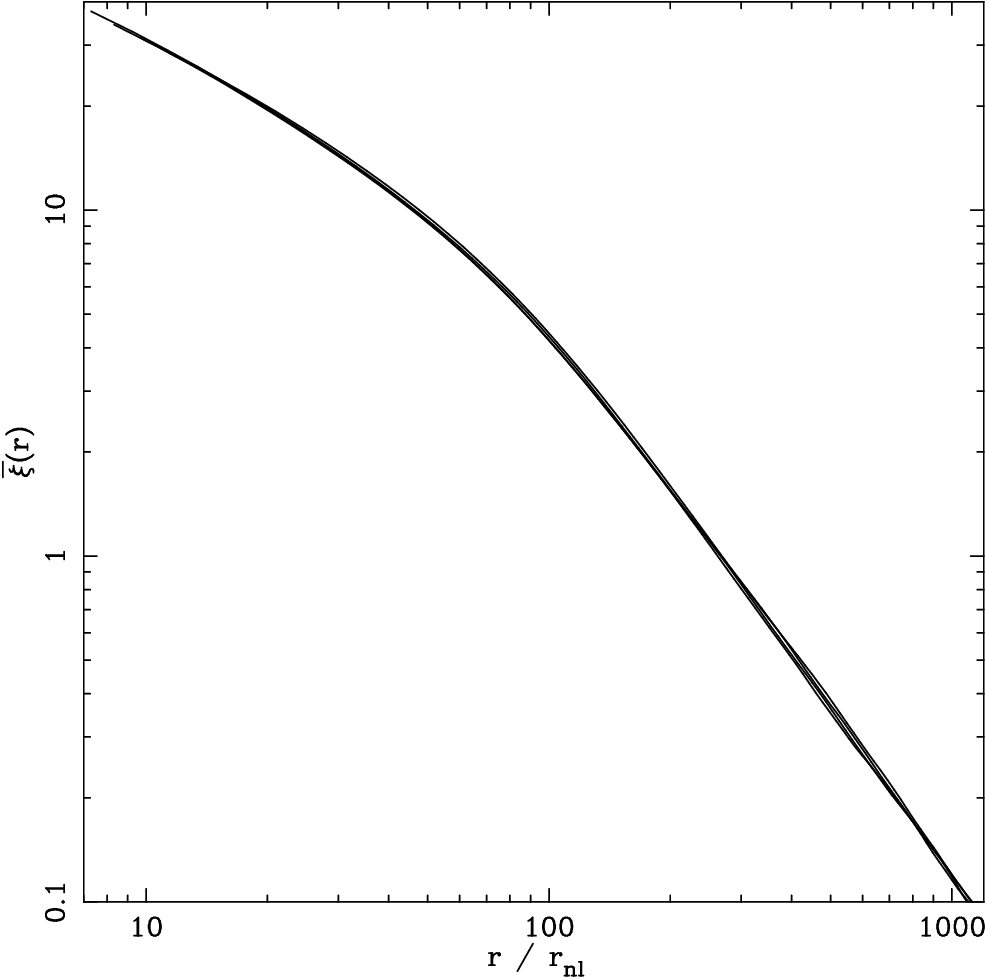}
\caption{This figure shows $\bar\xi$ as a function of
$r/r_{nl}(t)$ for several epochs.  Here $r_{nl}(t)$ is the scale which
is going non-linear at time $t$ and it varies in proportion with
$a^{2/\left(n+2\right)}$ in the Einstein deSitter model. The
index of the power spectrum is $n=-0.4$. We have only plotted 
$\bar\xi$ at scales more than two times larger than the artificial 
softening length used in the simulation.}
\label{fig_scale_inv}
\end{figure}

Fig.\ref{fig_scale_inv} shows $\bar\xi$ as a function of $r/r_{nl}(t)$ 
for several epochs obatined from a 2d TreePM simulation of a power law 
model with index $n=-0.4$. $r_{nl}(t)$ is the scale which is going 
non-linear at time $t$ and it varies in proportion with 
$a^{2/\left(n+2\right)}$ in the Einstein deSitter model. In the figure, 
we have only plotted $\bar\xi$ at scales more than two times larger 
than the artificial softening length used in the simulation. We can see 
that scale invariance holds for the spectrum over a wide range which 
means that we can probe the non-linear regime in gravitational 
clustering with a high degree of accuracy using the 2d TreePM code. 

\section{The 2d TreePM Code vs. the 2d Particle-Mesh Code}

In this section, we present a brief comparison of the 2d TreePM and 
Particle-Mesh methods with the aim of highlighting the efficacy of our
method in 2d cosmological simulations. 

We ran a 2d simulation of a power law model with index $n=-0.4$ 
with $1024^2$ particles on a $1024^2$ grid in an Einstein 
deSitter background Universe with a PM code as well as with the 
TreePM code discussed here. For the TreePM run we used $r_s=1$, 
$\theta_c=0.5$ and softening parameter $\epsilon = 0.2$.

\begin{figure}
\epsfxsize=2.46truein\epsfbox[42 28 518 510]{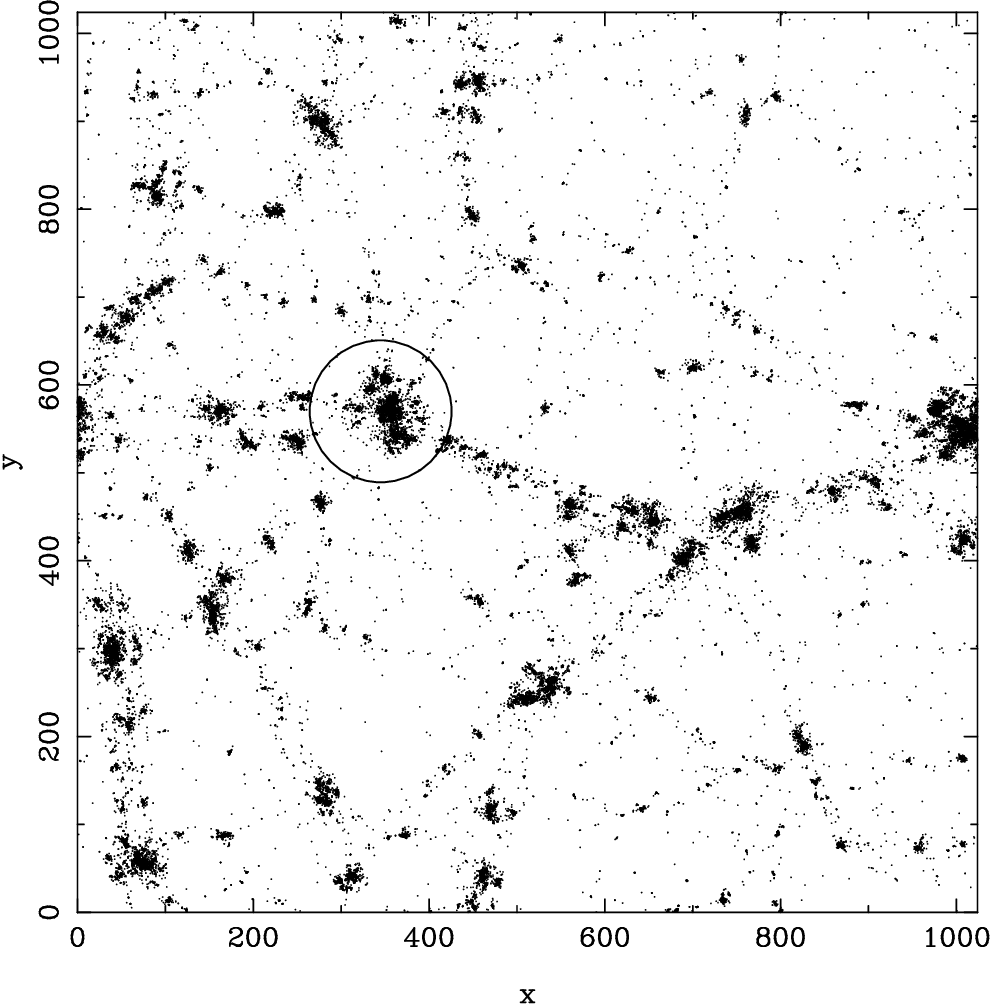}
\epsfxsize=2.46truein\epsfbox[42 28 518 510]{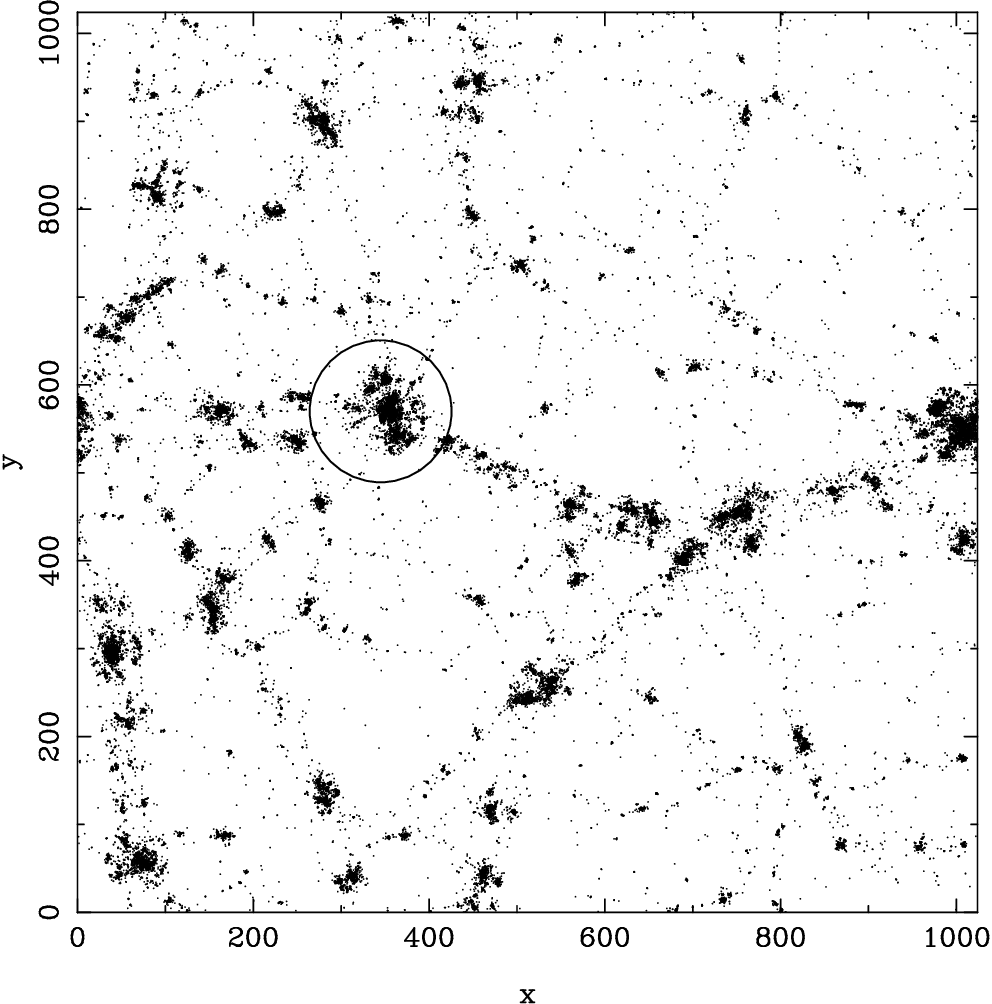}
\epsfxsize=2.46truein\epsfbox[42 28 518 510]{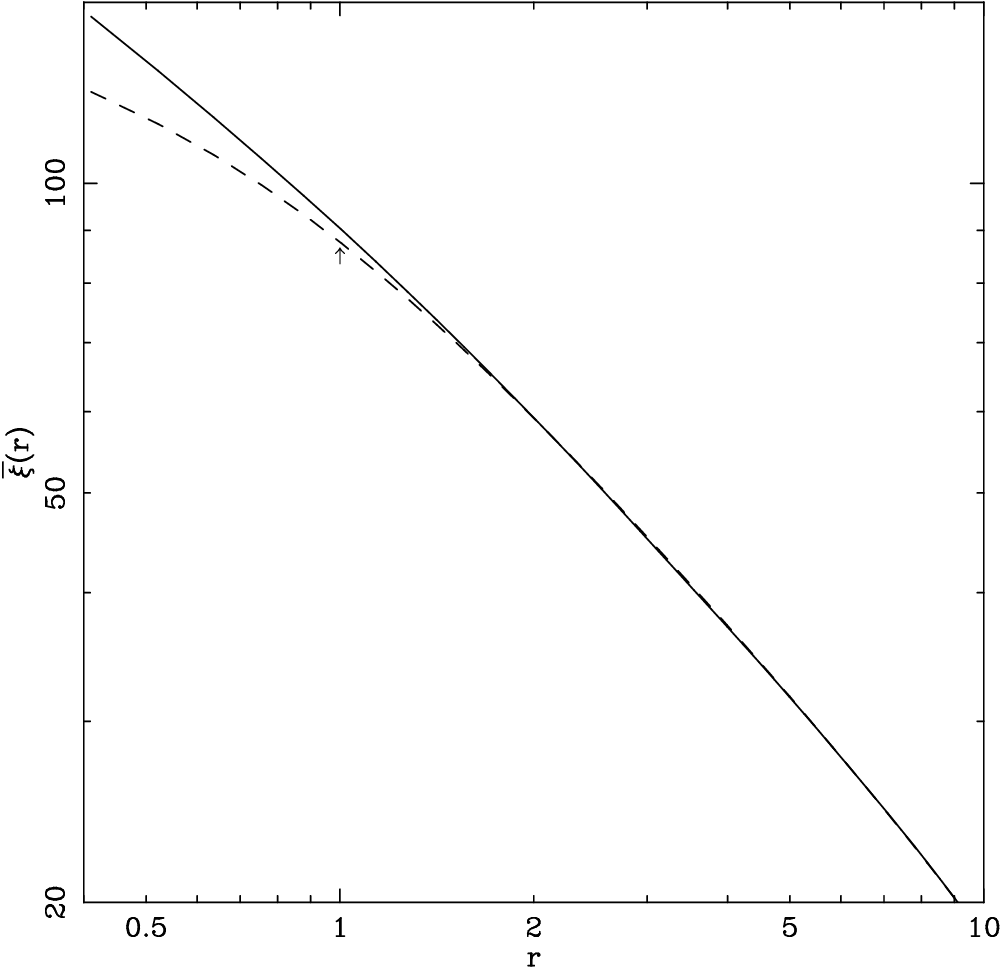}
\epsfxsize=2.46truein\epsfbox[42 28 518 510]{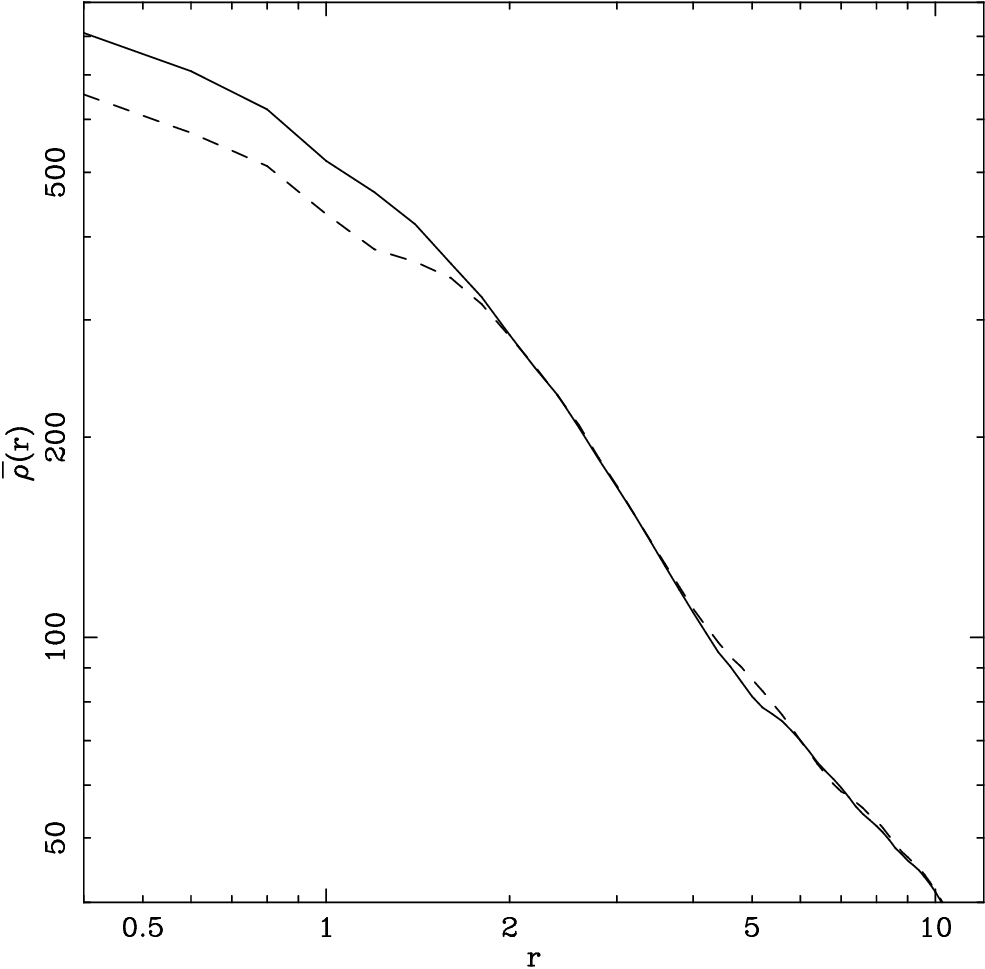}
\caption{The top panel of this figure shows a box from a 
simulation of a power law model with index $n=-0.4$ for two
cases. The top left panel shows the box from a TreePM simulation. 
For comparison, we have included the same box from a PM simulation 
of the same initial conditions in the top right panel. For convenience,
we have randomly plotted only one out of every sixteen particles in the 
original simulations in the figure. The large scale structures appear to 
be the same in the two. We will look more closely at the circled haloes 
in both panels in the following discussion. The bottom left panel of this 
figure shows the averaged correlation function $\bar\xi(r)$ as a function 
of scale in grid units. The thick line shows this quantity for the TreePM 
simulation and the dashed line shows the same for the PM simulation. We 
have only plotted $\bar\xi$ at scales more than two times larger than 
the artificial softening length used in the TreePM simulation. The point 
marked by an arrow in the figure represents the larger softening scale 
for the PM code. The correlation functions match at large scales but the 
PM simulation underestimates the clustering at small scales. The bottom 
right panel is a plot which shows average density $\bar\rho$ within a 
sphere of radius $r$ as a function of $r$ (in grid units) for the two 
halos circled in the top panel of this figure. Again, the thick line 
shows average density for the TreePM simulation and the dashed line 
shows the same for the PM simulation. Here also we have plotted average 
density at scales more than two times larger than the artificial softening 
length used in the TreePM simulation. The density profiles match at large 
scales as expected, but one can see that the TreePM simulation gives rise 
to haloes with higher central densities.}
\label{pos_den_corr_pm_treepm}
\end{figure}

The top panel of Fig.\ref{pos_den_corr_pm_treepm} shows identical boxes 
from two independent simulations with the same initial conditions. 
The top left panel shows a simulation with the TreePM code and the top 
right panel shows the same for a PM code. The large scale structures 
appear to be the same in the two. However, one can see that there are 
significant differences at small scales when one plots the two-point 
correlation function for the two cases. $\bar\xi(r)$ is plotted as a 
function of scale $r$ in the bottom left panel of 
fig.\ref{pos_den_corr_pm_treepm}. The thick line shows the correlation 
function for the TreePM simulation and the dashed line shows the same 
for the PM simulation. We have only plotted $\bar\xi$ at scales 
more than two times larger than the artificial softening length used in 
the TreePM simulation. The correlation function in the TreePM simulation 
matches with that from the PM simulation at large scales, but at scales 
of the order of unity (in grid units) and below, the TreePM simulation 
has a higher correlation function. This is to be expected because of the 
superior force resolution of the TreePM method as opposed to the PM force, 
where the force is softened naturally at the grid scale. The scale of 
softening for the PM code is marked by an arrow in the figure. 

We also study the density profiles of the two circled haloes 
in the top panel of fig.\ref{pos_den_corr_pm_treepm}. These particular 
haloes have been chosen as representatives for the analysis because 
they are reasonably large, dense and spherically symmetric. The bottom 
right panel of fig.\ref{pos_den_corr_pm_treepm} shows average 
density $\bar\rho$ within a sphere of radius $r$ from the halo centre 
plotted as a function of $r$ for the two haloes. The full line shows 
the density profile for the halo from the TreePM simulation and the 
dashed line the same from the PM simulation. Here we have only plotted 
average density at scales more than four times larger than the artificial 
softening length used in the TreePM simulation. We can see that, though 
not visibly obvious from fig.\ref{pos_den_corr_pm_treepm}, the halo 
from the TreePM simulation is clearly far denser in the central 
region as compared to the halo from the Particle-Mesh simulation. 
The density profiles converge at some distance from the halo 
centres as expected.

\section{Computational Requirements}

In this section, we describe the computational resources required for
the present implementation of the 2d TreePM code. Given that we have
combined the Tree and the PM codes, the memory requirement is obviously
greater than that for either one code. We need four arrays for the PM
part i.e. for the potential and the force. The rest is exactly the 
same as a standard Barnes and Hut 2d Tree code. With efficient memory 
management, we need less than $75$MB of RAM for a simulation with 
$1024^2$ particles on a $1024^2$ grid. The number mentioned is for 
floating point variables. If we use double precision variables, our 
requirement will go up by a factor of two.

The time taken (per time step per particle) by the 2d TreePM code
($r_s=1$, $\theta_c=0.5$, $r_{cut}=4.5 r_s$, $N_{particle}=1024^2$, 
$N_{grid}=1024^2$) is of the order of $240$ microseconds. This number 
was generated using a $2.4$GHz Xeon personal computer where the code 
was compiled with the Intel F90 compiler.

\section{Discussion}

In this paper, we have described the two-dimensional TreePM method in 
detail. Our method offers greater dynamic range and superior resolution 
as compared to a 2d Particle-Mesh method and can therefore probe the 
non-linear regime in two-dimensional cosmological simulations more 
effectively. We believe that a 2d TreePM code will allow us to 
explore a higher dynamic range in densities (and $\bar\xi$) 
for studying scaling relations in two-dimensions as compared to 
earlier work done using Particle-Mesh codes \cite{Bagla 1998}. Work
is in progress in this direction and will be reported elsewhere.

The 2d TreePM code is also amenable to parallelisation along the 
lines of the 3d TreePM code \cite{Bagla 2002b, Ray 2004} and is 
likely to scale well.

\section{Acknowledgement}

The work reported here was done using the Beowulf at
the Harish-Chandra Research Institute (http://cluster.mri.ernet.in).

I would sincerely like to thank J.S. Bagla for insightful discussions,
suggestions and comments.

\label{lastpage}

\end{document}